\documentclass[twocolumn, showpacs,preprintnumbers,amsmath,amssymb]{revtex4}
\usepackage{graphicx}
\usepackage{color}
\usepackage{epstopdf}
\usepackage{bm}
\newcommand{\figurewidth}{3in}

\begin{document}
\title{Confinement effects on the vibrational properties of III-V and II-VI nanoclusters}

\author{Peng Han}
\author{Gabriel Bester}
\affiliation{Max-Planck-Institut f\"ur Festk\"orperforschung, Heisenbergstr. 1, D-70569 Stuttgart, Germany}

\date{\today}
\begin{abstract}
We present a first-principles study of the confinement effects on the vibrational properties
of thousand atoms (radii up to 16.2~\AA)  colloidal III-V and II-VI nanoclusters.
We describe how the molecular-type vibrations, such as surface--optical, surface--acoustic and coherent acoustic modes, coexist and interact with bulk-type vibrations, such as longitudinal and transverse acoustic and optical modes. 
We link vibrational properties to structural changes induced by the surface and 
highlight the qualitative difference between III-Vs and II-VIs. We describe the size dependence of the vibrations and find good agreement for Raman shifts and for the frequency of coherent acoustic modes with experiments.  
\end{abstract}


\maketitle

Semiconductor nanocluster (NC), research is a rapidly growing field driven by the attractive idea to tailor material properties 
by acting on the morphology of the structures. The modification of the optical properties by merely changing the diameter 
of colloidal NCs (aka quantum dots) is one of the figurehead of nanostructure science. Indeed, the optical properties of NCs are well understood 
theoretically and well controlled experimentally. 
These experiments are usually performed at room temperature where vibrations are naturally involved. 
A solid understanding of the vibrational properties of NCs is a decisive step for the real world application of NCs, where the physical properties such as temperature broadening, loss of quantum coherence and relaxation of charge carriers are key components \cite{klimov00,pandey08,wijnen08,kilina09,huxter10}.
However, despite a rather intense recent experimental work on the vibrational properties of NCs \cite{klimov00,pandey08,krauss96,oron09,krauss97,chilla08}, 
the theoretical understanding of the processes involved remains rather shallow. 
Most of the work has been carried out at the level of continuum models (e.g. Ref. \cite{roca94}) or effective potentials (e.g. Ref. \cite{fu99,valentin08,han11}) that rely on parameters derived from the bulk. Only very recently, {\it ab initio} calculations of the vibrational properties have become possible for Silicon NCs with sizes comparable to the experimental situation\cite{khoo10}. Unexpected effects of competing bond-length contraction and undercoordination were unraveled\cite{khoo10}. In general
the study of this intermediate size regime, between bulk and molecule, is interesting not only from a technological point of view (how can we expect the properties to change?) but also from a fundamental point of view (how can molecular-type and bulk-type vibrations coexist?).

In this work, we study the vibrational properties of InP, InAs, GaP, GaAs, CdS, CdSe, 
and CdTe NCs with radii ranging from
10.7 (In$_{141}$P$_{140}$H$^*_{172}$) to 16.2~\AA~ (Cd$_{369}$S$_{348}$H$^*_{300}$) via \emph {ab initio} density functional theory (DFT). 
We find that structural relaxation effects are large and penetrate inside the core of the clusters for NCs up to radii of 15~\AA. 
These morphological changes are qualitatively different for III-Vs (where several surface layers contract inwards) and II-VIs (where long range ionic interactions lead to a large bond length distribution at the surface). These structural effects dominate the vibrational properties and lead to a large blue shift with decreasing cluster size in III-Vs and strong broadening of the optical branches and a filling of the optic-acoustic phonon gap with surface states in II-VIs. We find a small red shift of Raman peaks for our largest cluster sizes, in agreement with experiment. We analyze the bulk parentage of the cluster modes and find them to be fully mixed longitudinal-transverse and fully distinct acoustic-optical. We find two types of surface modes, the surface acoustic (SA) modes ---as the lowest frequency modes--- and the surface optical (SO) modes. We identify the {\it coherent acoustic phonons} and find good agreement with experiment and with results from the simple Lamb mo
 del. 

The NCs are constructed by cutting a sphere, 
centered on a cation, from the zinc blende bulk structure, and removing the surface atoms having only one
nearest neighbor bond. The surface dangling bonds are terminated by 
pseudo-hydrogen atoms H$^*$ with a fractional charge of $1/2$, $3/4$, $5/4$, and $3/2$ for group VI, V, III
and II atoms, respectively. The $T_d$ symmetry of the bulk is thereby conserved.

The calculations are performed using the local density approximation (LDA), Trouiller-Martin norm-conserving 
pseudopotentials, and an energy cut-off of 30 Ry for III-V and 40 Ry for II-VI clusters \cite{cpmd08}.
The forces are minimized to less than 3$\times$10$^{-6}$ Ha/Bohr
under constrained symmetry. The dynamical matrix elements are then calculated via finite difference
and the vibrational eigenmodes and eigenvectors are obtained after diagonalization. More details can be found in \cite{supplemental}.

To analyze the eigenmodes in terms of bulk and surface contributions, we calculate the 
projection coefficients
\begin{equation}
\label{eq:core}
\alpha^{\nu}_{c(s)}=\frac{\sum_{I}^{N_c (N_s)}|\bm X^{\nu}(I)|^2}{\sum_{I=1}^{N}|\bm X^{\nu}(I)|^2},
\end{equation}
where,  $N_c$, $N_s$, and $N$ are the core, surface, and total number of atoms, 
$\bm X^{\nu}(I)$ represents the three components that belong to atom $I$ 
from the $3N$-component eigenvector. 
The sum of $\alpha^{\nu}_{c}$ and $\alpha^{\nu}_{s}$
is equal to one. We define the surface atoms as the atoms belonging to the outermost seven layers of the cluster
(around 3~\AA~ thick).
The vibrational eigenmodes  are further projected onto bulk eigenmodes $\bm u_{n, \bm q}$ as
\begin{equation}
\label{eq:proj}
\bm X^{\nu}(I) = \sum_{n, \bm q} C_{n, \bm q}^{\nu}\bm U_{n, \bm q}(I),
\end{equation}
with
$\bm U_{n, \bm q}(I) = {\bm u^{\tau}_{n,{\bm q}}}\exp({i{\bm q}\cdot{\bm R_{I}}})$.
Here, ${\bm u^{\tau}_{n,{\bm q}}}$ are the three components of the 6-component bulk
eigenvectors corresponding to the atom of type $\tau$, 
phonon branch $n$ and  wave vector $\bm q$. $\bm {R_I}$ is the atom position of atom $I$ in the cluster.
The bulk phonon modes used in Eq.~(\ref{eq:proj}) are 
calculated via density functional perturbation theory \cite{abinit}.

\begin{figure}
\centerline{\includegraphics[width=\figurewidth]{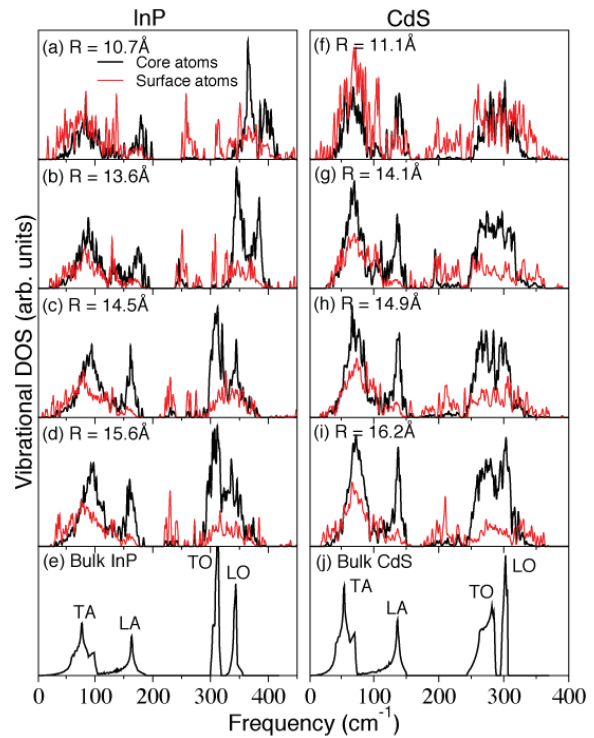}} \caption{
(Color online) Vibrational density of states (DOS) contributed by core atoms (black) and surface atoms
(red) for (a)-(d) InP clusters, (e) bulk InP, (f)-(i) CdS clusters and (j)
bulk CdS.
}\label{fig:vdos}
\end{figure}

We have calculated the vibrational properties for a total of 23 \cite{supplemental} clusters made of   
InP, InAs, GaP, GaAs, CdS, CdSe, and CdTe.  In this letter we extract the essence of 
these calculations and will select representative results.
The vibrational DOS of InP and CdS NCs and bulk are plotted with a broadening of 0.8 cm$^{-1}$ 
in Fig.~\ref{fig:vdos}.  
Although formally transverse- and longitudinal acoustic (TA, LA), and optical (TO, LO) modes seize to exist in a 
cluster, the comparison with the bulk Fig.~\ref{fig:vdos}(e)(j) reveals an obvious bulk  parentage.
At this stage, we make three observations: First, the III-V materials [illustrated by InP in Fig.\ref{fig:vdos}(a-d)] show a blue shift
of LO-, TO- and LA-derived cluster modes with decreasing size, 
while II-VI materials [illustrated by CdS in Fig.\ref{fig:vdos}(f-i)] show no such shifts. Second, the surface modes tend to 
completely fill the acoustic-optic phonon gap in II-VIs, while a gap remains for III-Vs. Third, the ``broadening" of the bulk 
optical phonon branches induced by the confinement is larger for II-VIs than for III-Vs. 

\begin{figure}
\centerline{\includegraphics[width=\figurewidth]{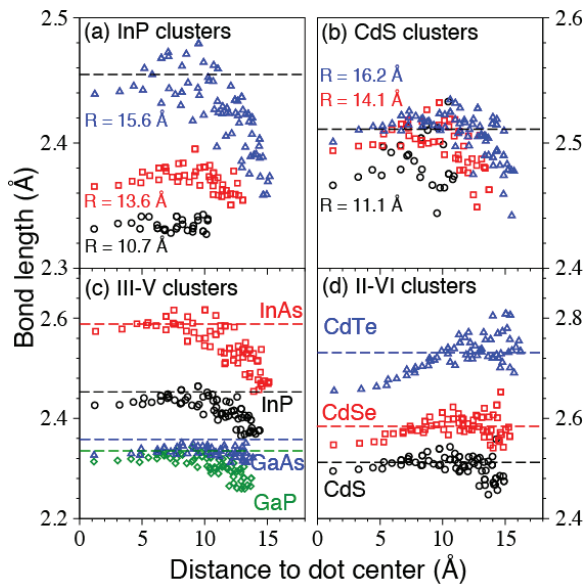}} \caption{
(Color online) Bond length distribution as a function of their distance to the dot center. For (a) InP, (b) CdS, (c) III-Vs and (d) II-VIs.
LDA bulk bond lengths are given as dashed lines.
}\label{fig:size_bond}
\end{figure}

In order to understand these three effects, we plot in Fig.~\ref{fig:size_bond} 
the nearest neighbor distances of relaxed III-Vs and II-VI NCs as a function of the 
distance of the respective bond to the cluster center. 
In all cases, the bond length at the dot center is reduced through the presence of the surface. For all the NCs studied, besides CdSe and CdTe, we find that  for R=15 {\AA} the bond length at the center has almost recovered and is 99\% of the bulk bond length. CdSe and especially CdTe recover their bulk bond length slower. 
The bond length distribution looks qualitatively different for the III-Vs and for the more ionic II-VI materials. For III-Vs the surface shells show a successive reduction of bond length, going outwards, while II-VIs show a large bond-length distribution, as previously observed for small clusters\cite{puzder04a}. The overall reduction of bond length in III-Vs along with the positive Gr\"{u}neisen parameters (describing the change in phonon frequencies with volume) explains the blue shift of the LO-, TO- and LA-derived cluster modes (first observation above). The lack of shift in the TA modes stems from the small negative Gr\"{u}neisen parameter for this branch. We attribute the  large bond-length distribution at the surface of II-VIs to the long-range ionic interaction and connect it to the broadening of optical branches and the filling of the phonon-gap (second and third observations above). 

\begin{figure}
\centerline{\includegraphics[width=\figurewidth]{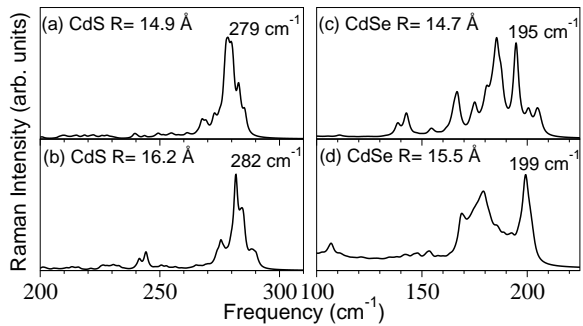}} \caption{
Calculated Raman spectra for CdS and CdSe nanocrystals for different sizes.}\label{fig:Raman}
\end{figure}

Experimentally,  the vibrational properties are mostly explored {\it via} Raman scattering. To compare our results with experiment we have calculated the Raman intensities in Fig. \ref{fig:Raman} using the theory proposed by Richter \emph{et al.} \cite{RICHTER81}, 
\begin{equation}
\label{eq:raman}
I(\omega)\propto \sum_{n,\nu,\bm{q}}\frac{|C_{n,\bm{q}}^{\nu}|^{2}}{(\omega - \omega^{\nu})^{2}+(\Gamma_0/2)^2},
\end{equation}
with our {\em ab initio} projection coefficient $C_{n,\bm{q}}^{\nu}$ and dot vibrational frequency $\omega^{\nu}$. For
the natural Lorentzian linewidth we use 2 cm$^{-1}$ and the coefficients $C_{n,\bm{q}}^{\nu}$ are summed up to $\Delta q \approx 1/(2R)$.
Although NCs with diameter down to 2 nm can be produced, Raman shifts have been reported for NCs with more than 3 nm diameter. Only our largest NCs reach this size and we focus in Fig. \ref{fig:Raman} on clusters with radii around 15 {\AA} and 16 {\AA}. 
For CdS and CdSe we obtain a small red shift of the most intense Raman peak with decreasing NC size. This is in agreement with experiments (CdS \cite{Verma00}, CdSe \cite{dzhagan09}), although experiments are performed on ensemble of NCs and a proper comparison would require an average over several NCs, allowing for small variations in shape and surface morphology. A nearly constant Raman peak \cite{guzelian96,Seong03} was reported for large InP clusters, which is consistent with our results for radii from 14.5 to 15.6~\AA~ in Fig.~\ref{fig:vdos} (c) and (d). We note that 4 meV blue shifts have been reported for InP (110) surfaces\cite{nienhaus95}.

\begin{figure}
\centerline{\includegraphics[width=\figurewidth]{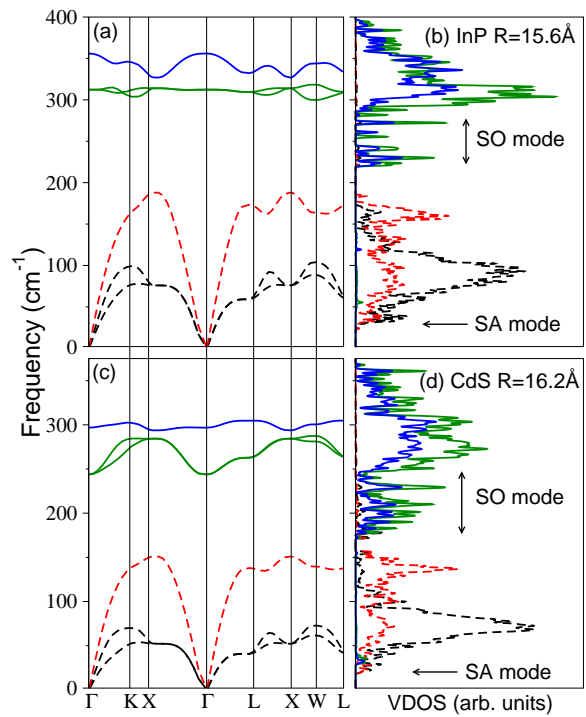}} \caption{
(Color online) Bulk phonon dispersion curves for (a) InP and (c) CdSe. Vibrational DOS projected onto the bulk modes for
(b) InP and (d) CdS NCs.
}\label{fig:proj}
\end{figure}

We now proceed to quantitatively analyze the bulk parentage of the cluster modes according to the projection in Eq.~(\ref{eq:proj}). 
From Fig.\ref{fig:vdos} we can see that besides some modes in the gap and at very low frequency, the modes are generally mixed bulk-surface (red and black) modes. This is true up to our largest clusters, although the surface character of ``bulk" modes has significantly decreased compared to the small clusters. 
In Fig.~\ref{fig:proj} (b) and (d), we show the vibrational DOS of InP and CdS clusters,
where the character (TA, LA, TO, and LO) of the modes are color coded (black, red,
green, and blue, respectively). The phonon dispersion
curves of bulk InP and CdS are plotted in Fig.~\ref{fig:proj} (a) and (c). From
Fig.~\ref{fig:proj} (b) and (d), we see that the cluster eigenmodes have either optical
or acoustic character with virtually no mixing. Interestingly, the ``pure'' surface modes
which vibrate with frequencies below the lowest core modes have purely acoustic 
character and the modes within the gap, between the acoustic and optical modes, have
purely optical character. These surface modes can be described as ``surface acoustic'' (SA) and ``surface optical'' (SO) modes,
respectively. 
We note that in principle, a mixing of acoustic and optical modes is allowed from symmetry 
arguments. Indeed, both acoustic and optical modes have the same symmetry within the $T_{d}$
point group\cite{Yu10}, and we can see small mixing around the frequency of SO  
modes in Fig.~\ref{fig:proj} (b) and (d).
From Fig.~\ref{fig:proj}, we also notice that the transverse and longitudinal
character of the mode gets lost in the cluster, which shows fully mixed 
transverse-longitudinal modes.
This mixing is introduced by the lack of translational symmetry. 
In NCs, the propagating waves with various wave vector $\bm q$
are combined into standing waves, pictorially via reflections at the boundary.
This result is consistent with the experimental measurements \cite{krauss96}, where Raman activity was measured
for frequencies around the TO-branch. 

\begin{figure}
\centerline{\includegraphics[width=\figurewidth]{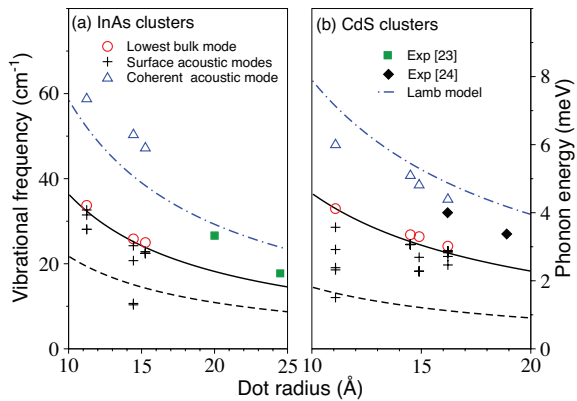}} 
\caption{(Color online) Lowest modes with bulk character (circles), surface acoustic modes (crosses), lowest breathing modes (triangles) and experimental results \cite{oron09,saviot98} for the coherent acoustic modes (square and diamonds). Lowest spheroidal mode according to the Lamb model (dashed dotted line), according to the confined bulk model using the sound velocity of the TA- (solid line) and LA-branch (dashed lines).}\label{fig:as}
\end{figure}

Besides the mixing of transverse and longitudinal characters, confinement has a direct effect on the 
lowest non-zero vibrational
frequency. Similarly to classical waves confined in one dimension, the
vibrational frequency of the longest wavelength in a NCs can be described as
$f_{min}=v/2d$. Here, $v$ is the velocity of sound and $d$ is the diameter of the cluster.
In Fig.~\ref{fig:as}, we plot the size-dependent lowest frequencies $f_{min}$
calculated from the longitudinal and transverse sound velocities as solid and dashed curves.
As circles and crosses we plot the lowest core and SA modes obtained from the DFT
calculations. We see that the lowest modes with dominant bulk character follow closely the analytic
$1/R$ dependence. 
The SA modes have the lowest frequency since they can run over the surface of the cluster. They are
however strongly affected by the morphology of the surface and their frequencies are not monotonous with cluster size.
One last important type of modes are the so called coherent acoustic modes. These are breathing modes, where all the atoms vibrate in phase. 
These modes have been observed with Raman spectroscopy, far-infrared absorption spectroscopy, and resonant high-resolution photoluminescence 
spectroscopy\cite{krauss97,saviot98,chilla08,oron09},
and are now the center of attention when the manipulation of spins and the spin-decoherence is investigated.
We plot our results as triangles in Fig.\ref{fig:as}, the experimental results as diamonds and the results of the 
Lamb model\cite{lamb82} as dashed dotted line. Our results are in good agreement with the experimental results (although our clusters sizes are still somewhat smaller than experiment in the case of InAs) and with the simple Lamb model. In the continuum Lamb model, only the bulk transverse and longitudinal sound velocity enters. It may therefore be surprising to give such a good description, although we described that the clusters  undergo strong lattice relaxations, absent in a continuum description. We attribute the validity of Lamb model to  the small dependence of the sound velocities on the bond-length (2\% change for a lattice deformation of 0.1~\AA) and the collective nature of the modes, where an averaging of the lattice deformations (especially for II-VIs) is a good description. 

By studying the vibrational properties of III-V and II-VI NCs with up to one thousand atoms  via 
\emph{ab initio} DFT we identify the following confinement effects. The LA, TO and LO-derived 
cluster modes of  III-V clusters significantly blue shift with decreasing cluster size. For II-VI clusters this shift is 
absent but the broadening of bulk derived modes is significant and the gap between optical and 
acoustic phonons is filled by surface modes.  We can clearly ascribe these observations to the large
relaxation of the clusters dominated by: an inward relaxation of the surface penetrating deep inside the cluster in case of the III-Vs and
a large distribution of bond length at the surface of II-VIs. These strong confinement effects tend to disappear for clusters with more 
than 1000 atoms, where a small red shift of the Raman peaks (we can see for our largest clusters) remains, due to a softening in response to undercoordination. The modes are all fully transverse-longitudinal mixed but surprisingly distinct in terms of their optical acoustic characters. We find surface optical modes in the phonon gap and surface acoustic modes as the lowest frequency modes. The {\it coherent acoustic phonons} are identified and found to be in good agreement with results from the Lamb model and experiment. We explain why the simple model by Lamb gives an accurate description in case of the breathing modes while the vibrational properties of small NCs are poorly described by continuum models in general. 

\begin{acknowledgements}
We acknowledge financial support by the Marie Curie Reintegration Grant. Most of the simulations were preformed on the national supercomputer NEC Nehalem Cluster at the High Performance Computing Center Stuttgart (HLRS).
 \end{acknowledgements}


\end{document}